\documentclass[aps,pra,twocolumn,showpacs,preprintnumbers]{revtex4-2}
\usepackage{amsmath,amssymb,bm}
\usepackage[T1]{fontenc}
\usepackage[utf8]{inputenc}
\usepackage{lmodern}
\usepackage{amsfonts,mathtools}
\usepackage{geometry}
\usepackage{microtype}
\usepackage{hyperref}
\newcommand{\trace}{\operatorname{Tr}}
\newcommand{\dif}{\mathrm{d}}
\newcommand{\ii}{\mathrm{i}}

\newcommand{\cre}[1]{a_{#1}^{\dagger}}
\newcommand{\ann}[1]{a_{#1}}
\newcommand{\commutator}[2]{\left[#1,#2\right]}
\newcommand{\anticomm}[2]{\left[#1,#2\right]_{+}}
\newcommand{\expect}[1]{\langle 0|#1|0\rangle}
\newcommand{\ket}[1]{\left|#1\right\rangle}
\newcommand{\bra}[1]{\left\langle #1\right|}
\newcommand{\texpect}[1]{\langle \tilde 0|#1|\tilde 0\rangle}
\newcommand{\mat}[1]{\mathbf{#1}}

\begin{document}

\title{A density-matrix derivation of the Hartree--Fock equations in a nonorthogonal atomic-orbital basis}

\author{Thomas Kjærgaard}
\affiliation{Kvantify ApS, DK-2100 Copenhagen Ø, Denmark}

\date{\today}

\begin{abstract}
We present a pedagogical derivation of the Hartree--Fock equations using the
second-quantization atomic-orbital density-matrix formalism developed by
Kjærgaard, Jørgensen, Olsen, Coriani, and Helgaker for AO-based response
theory.  The purpose is to introduce an alternative derivation of the Hartree--Fock equation, showing that the standard AO Hartree--Fock stationarity condition follows naturally
from the exponential parametrization of the one-particle density matrix in a
nonorthogonal AO basis.  This route provides a compact bridge between elementary
Hartree--Fock theory and the density-matrix machinery used in modern response
theory and linear-scaling formulations.
\end{abstract}

\maketitle

\section{Introduction}

The Hartree--Fock equations are most commonly derived by minimizing the
single-determinant energy with respect to molecular-orbital coefficients subject
to orthonormality constraints.  This route leads directly to the familiar
Roothaan--Hall equations,
\begin{equation}
  \mat F\mat C = \mat S\mat C\bm\varepsilon,
\end{equation}
and is therefore well suited for introducing self-consistent-field theory~\cite{Roothaan1951,Hall1951,SzaboOstlund,Helgaker2000}~.
 In this formulation, the molecular orbitals are the primary variational objects, while
the density matrix is usually introduced afterward as a convenient compact
representation of the occupied orbital space.

An alternative viewpoint is to regard the one-particle density matrix itself as the fundamental object. Density-matrix formulations of many-electron theory go back to the classic work of L\"owdin and McWeeny, where idempotency, particle number, and representability conditions were emphasized as defining properties of a single-determinant state~\cite{Lowdin1955,McWeeny1960}.  In a
nonorthogonal atomic-orbital basis, these conditions acquire explicit metric
factors, and the usual idempotency condition becomes
\begin{equation}
  \mat D\mat S\mat D = \mat D .
\end{equation}
An AO-based density-matrix viewpoint is especially useful in response
theory. Larsen, J{\o}rgensen, Olsen, and Helgaker formulated Hartree--Fock and Kohn--Sham time-dependent response theory directly in an atomic-orbital basis, using an exponential parametrization of the density matrix to generate metric-preserving variations~\cite{Larsen2000}. This approach was further
developed by Kj{\ae}rgaard, J{\o}rgensen, Olsen, Coriani, and Helgaker in a second-quantization atomic-orbital formalism suitable for 
linear-scaling Hartree--Fock and Kohn--Sham response theory~\cite{Kjaergaard2008}. In this formalism, one works directly with creation and annihilation operators associated with nonorthogonal atomic orbitals. The metric of the AO basis enters the anticommutation relations, and the one-particle density matrix is transformed by an exponential parametrization that preserves particle number and idempotency.

The present note reuses this second-quantization AO formalism for a more
elementary purpose: to derive the ordinary ground-state Hartree--Fock equations.
Most of the theoretical machinery is therefore reproduced from
Ref.~\cite{Kjaergaard2008}; the pedagogical point made here is that the same
density-matrix parametrization gives a compact alternative derivation of the
basic Hartree--Fock stationarity condition,
\begin{equation}
  \mat F\mat D\mat S - \mat S\mat D\mat F = 0.
\end{equation}
The Roothaan--Hall equation is then recovered as a corollary.

Throughout, we assume a finite linearly independent AO basis and real orbitals.  With this assumption, all orbital coefficients, density matrices, and integrals are real.

\section*{Notation}
Greek indices label nonorthogonal atomic spin orbitals (AOs), while Roman indices label orthonormal molecular spin orbitals.  The AO metric is denoted by \(\mat S\), and the AO creation and annihilation operators are \(\cre\mu\) and \(\ann\mu\).  The expectation-value matrix
\begin{equation}
  \Delta_{\mu\nu}=\expect{\cre\mu\ann\nu}
\end{equation}
is distinguished from the standard AO density matrix \(\mat D\). 

\section{Second-Quantization-Based AO Theory}

\subsection{The density matrix}
Consider a set of real nonorthogonal atomic spin orbitals with real symmetric metric \(\mat S\).  The AO creation and annihilation operators satisfy
\begin{subequations}
\begin{align}
  \anticomm{\cre\mu}{\cre\nu} &= 0, \\
  \anticomm{\ann\mu}{\ann\nu} &= 0, \\
  \anticomm{\cre\mu}{\ann\nu} &= S_{\nu\mu}.
\end{align}
\end{subequations}
For a single-determinant reference state \(\ket{0}\), the expectation values of the creation and annihilation operators are given by
\begin{equation}
  \Delta_{\mu\nu}=\expect{\cre\mu\ann\nu}.
\end{equation}
where the determinant is
\begin{equation}
  \ket{0}=\cre i\cre j\cdots \cre l\ket{\mathrm{vac}},
\end{equation}
where $\cre i\cre j\cdots \cre l$ refer to the set of orthonormal molecular spin orbitals that are occupied in \(\ket{0}\). Roman letters are used here for the orthonormal molecular spin orbitals and Greek letters for their atomic counterparts.

The molecular spin orbitals are expanded in the AO basis as
\begin{equation}
  \cre i=\sum_\mu C_{\mu i}\cre\mu.
\end{equation}
The standard AO density matrix is
\begin{equation}
  D_{\mu\nu}=\sum_{i=1}^{n} C_{\mu i} C_{\nu i},
\end{equation}
where the sum runs over occupied spin orbitals, and the MO coefficient matrix is normalized according to
\begin{equation}
  \mat C^{T}\mat S\mat C=\mat I.
\end{equation}
The expectation-value matrix \(\bm\Delta\) and the standard AO density matrix \(\mat D\) are related by
\begin{equation}
  \bm\Delta = \mat S^{T}\mat D^{T}\mat S^{T}. \label{eq:delta-d-relation1}
\end{equation}
To prove this relation, use
\begin{subequations}
\begin{align}
  \anticomm{\cre p}{\ann\nu} &= \sum_\mu C_{\mu p}S_{\nu\mu}, \\
  \anticomm{\cre\mu}{\ann p} &= \sum_\nu C_{\nu p}S_{\nu\mu}.
\end{align}
\end{subequations}
It follows that
\begin{align}
&\bra{\mathrm{vac}}\ann l\cdots \ann j\ann i\cre\mu\ann\nu\cre i\cre j\cdots\cre l\ket{\mathrm{vac}} \nonumber\\
&\quad = \bra{\mathrm{vac}}\ann l\cdots \ann j\cre\mu\ann\nu\cre j\cdots\cre l\ket{\mathrm{vac}}\\
 &\qquad + \sum_{\lambda\sigma} C_{\lambda i}S_{\nu\lambda}C_{\sigma i}S_{\sigma\mu}.
\end{align}
Repeated use of this identity shows that Eq.~\eqref{eq:delta-d-relation1} is valid. Hence the AO density matrix element \(D_{\mu\nu}\) is identical to the expectation value \(\Delta_{\mu\nu}\) only in an orthonormal AO basis.

For two-electron expectation values, we define
\begin{equation}
  \Gamma_{\mu\nu\lambda\sigma}
  =\expect{\cre\mu\cre\lambda\ann\sigma\ann\nu}.
\end{equation}
which can be shown to factorizes as
\begin{equation}
  \Gamma_{\mu\nu\lambda\sigma}
  =\Delta_{\mu\nu}\Delta_{\lambda\sigma}-\Delta_{\mu\sigma}\Delta_{\lambda\nu}. \label{eq:wick-factorization}
\end{equation}
This is the AO analogue of the usual factorization of the two-electron density matrix in an orthonormal molecular spin-orbital basis.  In the standard nomenclature, however, \(\mat D\) is called the AO density matrix, while \(\bm\Delta\) and \(\bm\Gamma\) are expectation-value matrices of one- and two-electron operators.

The usual single-determinant density matrix satisfies
\begin{subequations}
\begin{align}
  \mat D^{T} &= \mat D, \\
  \trace(\mat D\mat S) &= N_{\mathrm{el}}, \\
  \mat D\mat S\mat D &= \mat D.
\end{align}
\end{subequations}
Using Eq.~\eqref{eq:delta-d-relation1}, these conditions become
\begin{subequations}
\begin{align}
  \bm\Delta^{T} &= \bm\Delta, \\
  \trace(\bm\Delta\mat S^{-1}) &= N_{\mathrm{el}}, \\
  \bm\Delta\mat S^{-1}\bm\Delta &= \bm\Delta.
\end{align}
\end{subequations}
The two sets of conditions are formally equivalent.  The \(\mat D\)-based form is usually simpler because it contains \(\mat S\), whereas the \(\bm\Delta\)-based form contains \(\mat S^{-1}\).  These relations are necessary and sufficient for a density matrix to represent a normalized single-determinant wave function.

\subsection{Transformations of the density matrix}
We next consider how \(\bm\Delta\) changes when the reference determinant is transformed by
\begin{equation}
  \hat T = \exp(\ii\hat\kappa),
\end{equation}
where \(\hat\kappa\) is a real symmetric one-electron operator,
\begin{equation}
  \hat\kappa = \sum_{\mu\nu}\kappa_{\mu\nu}\cre\mu\ann\nu,
\end{equation}
and \(\bm\kappa\) is a real symmetric matrix.  The transformed determinant is
\begin{align}
  \ket{\tilde 0}&=\exp(\ii\hat\kappa)\ket{0}
  =\exp(\ii\hat\kappa)\cre i\cre j\cdots\cre l\ket{\mathrm{vac}}\\
  &=\tilde a_i^{\dagger}\tilde a_j^{\dagger}\cdots\tilde a_l^{\dagger}\ket{\mathrm{vac}},
\end{align}
where we have introduced the transformed creation operators
\begin{equation}
  \tilde a_\mu^{\dagger}
  = \exp(\ii\hat\kappa)\cre\mu\exp(-\ii\hat\kappa).
\end{equation}
They obey the same AO anticommutation relations as the original operators:
\begin{align}
  \anticomm{\tilde a_\mu^{\dagger}}{\tilde a_\nu}
  &= \anticomm{\exp(\ii\hat\kappa)\cre\mu\exp(-\ii\hat\kappa)}{\exp(\ii\hat\kappa)\ann\nu\exp(-\ii\hat\kappa)} \nonumber\\
  &= \exp(\ii\hat\kappa)\anticomm{\cre\mu}{\ann\nu}\exp(-\ii\hat\kappa)
   = S_{\nu\mu}.
\end{align}
Thus the exponential transformations conserve the AO metric.

Using the Hausdorff expansion and the AO anticommutation relations, the transformed creation operator becomes
\begin{align}
  \tilde a_\mu^{\dagger}
  &= \cre\mu + \ii\commutator{\hat\kappa}{\cre\mu}
     -\frac{1}{2}\commutator{\hat\kappa}{\commutator{\hat\kappa}{\cre\mu}}+\cdots \nonumber\\
  &= \cre\mu
   + \ii\sum_\nu (\bm\kappa\mat S)_{\mu\nu}\cre\nu
   - \frac{1}{2}\sum_\nu (\bm\kappa\mat S)^2_{\mu\nu}\cre\nu +\cdots \nonumber\\
  &= \sum_\nu \left[\exp(\ii\bm\kappa\mat S)\right]_{\mu\nu}\cre\nu .
\end{align}
For \(\mat S=\mat I\), this reduces to an ordinary unitary transformation of the orbitals.

The transformed one-electron expectation-value matrix becomes
\begin{align}
  \tilde\Delta_{\mu\nu}
  &=\texpect{\cre\mu\ann\nu} \nonumber\\
  &=\bra{0}\exp(-\ii\hat\kappa)\cre\mu\exp(\ii\hat\kappa)
        \exp(-\ii\hat\kappa)\ann\nu\exp(\ii\hat\kappa)\ket{0}.
\end{align}
The inverse transformations are
\begin{subequations}
\begin{align}
  \exp(-\ii\hat\kappa)\cre\mu\exp(\ii\hat\kappa)
  &= \sum_\sigma \left[\exp(-\ii\bm\kappa\mat S)\right]_{\sigma\mu}\cre\sigma, \\
  \exp(-\ii\hat\kappa)\ann\nu\exp(\ii\hat\kappa)
  &= \sum_\sigma \left[\exp(\ii\mat S\bm\kappa)\right]_{\nu\sigma}\ann\sigma.
\end{align}
\end{subequations}
Consequently,
\begin{equation}
  \tilde{\bm\Delta}
  = \exp(-\ii\mat S^{T}\bm\kappa^{T})\,\bm\Delta\,
    \exp(\ii\bm\kappa^{T}\mat S^{T}). \label{eq:delta-transform}
\end{equation}
If \(\bm\Delta\) satisfies the Hermiticity, trace, and idempotency conditions above, then \(\tilde{\bm\Delta}\) satisfies them as well.  The transformed state is therefore again a normalized single determinant.

All matrices satisfying the single-determinant density-matrix conditions can be generated by an appropriate \(\bm\kappa\).  To remove redundant parameters, \(\bm\kappa\) may be projected as
\begin{align}
  \bm\kappa &= P(\bm\kappa)=\mat P\bm\kappa\mat Q^{T}+\mat Q\bm\kappa\mat P^{T},\\
  \mat P&=\mat D\mat S,\\
  \mat Q&=\mat I-\mat D\mat S.
\end{align}
This projection removes the redundant occupied--occupied and virtual--virtual blocks of the orbital-rotation parameter.  These redundant rotations change only the representation of the occupied or virtual subspaces and therefore leave the density matrix unchanged.  The non-redundant part of $\bm\kappa$ is the occupied--virtual plus virtual--occupied block, which generates genuine variations of the determinant.

Thus this projection eliminate the set of parameters that leave \(\bm\Delta\) unchanged \(\tilde{\bm\Delta}(\bm\kappa)=\bm\Delta\).

\section{The Hamiltonian}

In a nonorthogonal AO basis, the unperturbed Hamiltonian can be written as
\begin{align}
  \hat H_0
  &=\sum_{\mu\nu}
  \left(\mat S^{-1}\mat h\mat S^{-1}\right)_{\mu\nu}
  \cre\mu\ann\nu \nonumber\\
  &\quad +\frac{1}{2}\sum_{\mu\nu\kappa\tau}
  \sum_{\alpha\beta\gamma\delta}
  (S^{-1})_{\mu\alpha}(S^{-1})_{\beta\nu}
  (S^{-1})_{\kappa\gamma}(S^{-1})_{\delta\tau}\\
  &\quad \times g_{\alpha\beta\gamma\delta}
  \cre\mu\cre\kappa\ann\tau\ann\nu .
  \label{eq:h0-nonorthogonal-expanded}
\end{align}
Here the one- and two-electron integrals are
\begin{equation}
  h_{\alpha\beta}
  =\int \chi_{\alpha}(\mathbf r)
  \left(-\frac{1}{2}\nabla^2-\sum_A\frac{Z_A}{|\mathbf R_A-\mathbf r|}\right)
  \chi_{\beta}(\mathbf r)\,\dif\mathbf r,
\end{equation}
\begin{equation}
  g_{\alpha\beta\gamma\delta}
  =\iint
  \frac{\chi_{\alpha}(\mathbf r_1)\chi_{\gamma}(\mathbf r_2)
        \chi_{\beta}(\mathbf r_1)\chi_{\delta}(\mathbf r_2)}{r_{12}}
  \,\dif\mathbf r_1\,\dif\mathbf r_2 .
\end{equation}
Although the Hamiltonian itself contains inverse metric factors, these factors disappear from the final working expression.

\subsection{The ground-state energy}
The Hartree--Fock energy is the expectation value of the Hamiltonian with respect to the single-determinant state \(\ket{0}\).  Evaluating this using the one- and two-electron expectation-value matrices and the relation \(\bm\Delta = \mat S^T \mat D^T \mat S^T\), one obtains
\begin{align}
  E &= \expect{\hat H} \nonumber\\
    &= \sum_{\mu\nu}\left(\mat S^{-1}\mat h\mat S^{-1}\right)_{\mu\nu}\Delta_{\mu\nu}\\
     &+ \frac{1}{2}\sum_{\mu\nu\kappa\tau}
       \left(\mat S^{-1}\mat g\mat S^{-1}\right)_{\mu\nu\kappa\tau}
       \Gamma_{\mu\nu\kappa\tau}.
       \label{eq:energy-expectation-raw}
\end{align}
Using the factorization of Eq.~\eqref{eq:wick-factorization} and substituting \(\bm\Delta = \mat S^T\mat D^T\mat S^T\), all inverse metric factors cancel and the energy reduces to the familiar form
\begin{equation}
  E = \trace(\mat h\mat D) + \frac{1}{2}\trace\bigl(\mat G(\mat D)\mat D\bigr),
  \label{eq:hf-energy}
\end{equation}
where the Coulomb-exchange matrix is
\begin{equation}
  G_{\mu\nu}(\mat D)
  =\sum_{\lambda\sigma}D_{\lambda\sigma}
  \left(g_{\mu\nu\lambda\sigma}-g_{\mu\sigma\lambda\nu}\right).
  \label{eq:g-functional}
\end{equation}
The Fock matrix is defined as
\begin{equation}
  F_{\mu\nu} = \frac{\partial E}{\partial D_{\nu\mu}} = h_{\mu\nu} + G_{\mu\nu}(\mat D).
  \label{eq:fock-matrix}
\end{equation}

\subsection{Derivation of the Hartree--Fock equations}
We now derive the Hartree--Fock equations by requiring the energy to be stationary with respect to the orbital-rotation parameters \(\bm\kappa\).  Using the transformation in Eq.~\eqref{eq:delta-transform}, the corresponding transformed density matrix is
\begin{align}
  \tilde{\mat D} &= \mat S^{-1}\tilde{\bm\Delta}^T\mat S^{-1}\\
  &= \exp(\ii\bm\kappa\mat S)\,\mat D\,\exp(-\ii\mat S\bm\kappa).
  \label{eq:d-transformed}
\end{align}
Expanding to first order in \(\bm\kappa\),
\begin{equation}
  \tilde{\mat D} = \mat D + \ii\bm\kappa\mat S\mat D - \ii\mat D\mat S\bm\kappa + O(\kappa^2).
  \label{eq:d-first-order}
\end{equation}
The first-order change in the density matrix is therefore
\begin{equation}
  \mat D^{(1)} = \ii(\bm\kappa\mat S\mat D - \mat D\mat S\bm\kappa).
  \label{eq:d1}
\end{equation}

The energy of the transformed state is
\begin{equation}
  E(\bm\kappa) = \trace\bigl(\mat h\tilde{\mat D}\bigr) + \frac{1}{2}\trace\bigl(\mat G(\tilde{\mat D})\tilde{\mat D}\bigr).
  \label{eq:energy-kappa}
\end{equation}
We now substitute the expansion of \(\tilde{\mat D}\) from Eq.~\eqref{eq:d-first-order} into Eq.~\eqref{eq:energy-kappa} and collect terms order by order in \(\bm\kappa\).  Consider first the one-electron contribution:
\begin{align}
  \trace\bigl(\mat h\tilde{\mat D}\bigr)
  &= \trace\bigl(\mat h\mat D\bigr)
   + \ii\,\trace\bigl(\mat h(\bm\kappa\mat S\mat D - \mat D\mat S\bm\kappa)\bigr)
   + O(\kappa^2).
   \label{eq:one-electron-expand}
\end{align}
For the two-electron contribution, we use the linearity of \(\mat G\) in the density matrix, \(\mat G(\mat D + \mat D^{(1)}) = \mat G(\mat D) + \mat G(\mat D^{(1)})\), to obtain
\begin{align}
  &\frac{1}{2}\trace\bigl(\mat G(\tilde{\mat D})\tilde{\mat D}\bigr) \nonumber\\
  &\quad= \frac{1}{2}\trace\bigl(\mat G(\mat D)\mat D\bigr)
   + \frac{1}{2}\trace\bigl(\mat G(\mat D^{(1)})\mat D\bigr)\\
   &\quad+ \frac{1}{2}\trace\bigl(\mat G(\mat D)\mat D^{(1)}\bigr)
   + O(\kappa^2) \nonumber\\
  &\quad= \frac{1}{2}\trace\bigl(\mat G(\mat D)\mat D\bigr)
   + \trace\bigl(\mat G(\mat D)\mat D^{(1)}\bigr)\\
   &\quad + O(\kappa^2),
   \label{eq:two-electron-expand}
\end{align}
where in the last step we used the symmetry property \(\trace(\mat G(\mat A)\mat B) = \trace(\mat G(\mat B)\mat A)\).  Combining Eqs.~\eqref{eq:one-electron-expand} and \eqref{eq:two-electron-expand}, and substituting \(\mat D^{(1)}\) from Eq.~\eqref{eq:d1}, the energy becomes
\begin{align}
  E(\bm\kappa)
  &= \trace\bigl(\mat h\mat D\bigr) + \frac{1}{2}\trace\bigl(\mat G(\mat D)\mat D\bigr) \nonumber\\
  &\quad + \ii\,\trace\bigl(\bigl[\mat h + \mat G(\mat D)\bigr]
    (\bm\kappa\mat S\mat D - \mat D\mat S\bm\kappa)\bigr) + O(\kappa^2).
  \label{eq:energy-kappa-combined}
\end{align}
We now define the Fock matrix
\begin{equation}
  \mat F = \mat h + \mat G(\mat D),
  \label{eq:fock-definition}
\end{equation}
so that Eq.~\eqref{eq:energy-kappa-combined} takes the compact form
\begin{align}
  E(\bm\kappa)
  &= E(0) + \ii\,\trace\bigl(\mat F\bm\kappa\mat S\mat D - \mat F\mat D\mat S\bm\kappa\bigr) + O(\kappa^2) \nonumber\\
  &= E(0) + \ii\,\trace\bigl(\bm\kappa(\mat S\mat D\mat F - \mat F\mat D\mat S)\bigr) + O(\kappa^2),
  \label{eq:energy-kappa-expansion}
\end{align}
where in the last step we used the cyclic invariance of the trace.  The stationarity condition is
\begin{equation}
  \frac{\partial E(\bm\kappa)}{\partial \kappa_{\mu\nu}}\bigg|_{\bm\kappa=0}
  = \ii\left(\mat S\mat D\mat F - \mat F\mat D\mat S\right)_{\nu\mu} = 0.
  \label{eq:stationarity}
\end{equation}
Since this must hold for all \(\mu\nu\), we require
\begin{equation}
  \boxed{\mat F\mat D\mat S - \mat S\mat D\mat F = 0.}
  \label{eq:hf-ao}
\end{equation}
This is the Hartree--Fock equation expressed entirely in the atomic-orbital basis. 

\subsection{Equivalence to the Roothaan--Hall equations}
To establish the connection to the standard form, expand the density matrix as \(\mat D = \mat C_\mathrm{occ}\mat C_\mathrm{occ}^T\) and multiply Eq.~\eqref{eq:hf-ao} from the right by \(\mat C_\mathrm{occ}\).  Using the orthonormality condition \(\mat C^T\mat S\mat C = \mat I\), one obtains
\begin{equation}
  \mat F\mat C_\mathrm{occ} = \mat S\mat C_\mathrm{occ}\bm\varepsilon,
  \label{eq:roothaan-hall}
\end{equation}
where \(\bm\varepsilon = \mat C_\mathrm{occ}^T\mat F\mat C_\mathrm{occ}\) is the matrix of orbital energies (which can always be chosen diagonal by an orthogonal transformation among the occupied orbitals).  Equation~\eqref{eq:roothaan-hall} is the Roothaan--Hall eigenvalue equation.

Conversely, suppose that the occupied orbitals form an invariant subspace of the
generalized Fock problem,
\begin{equation}
  \mat F\mat C_\mathrm{occ}
  =
  \mat S\mat C_\mathrm{occ}\bm\varepsilon_\mathrm{occ},
\end{equation}
with
\begin{equation}
  \mat C_\mathrm{occ}^{T}\mat S\mat C_\mathrm{occ}=\mat I .
\end{equation}
The corresponding density matrix
\begin{equation}
  \mat D=\mat C_\mathrm{occ}\mat C_\mathrm{occ}^{T}
\end{equation}
then satisfies
\begin{equation}
  \mat D\mat S\mat D=\mat D,
\end{equation}
and projects onto the occupied subspace in the AO metric.  Multiplying the
generalized eigenvalue equation by \(\mat C_\mathrm{occ}^{T}\) from the
right gives
\begin{equation}
  \mat F\mat D
  = \mat S
  \mat C_\mathrm{occ}\bm\varepsilon_\mathrm{occ}
  \mat C_\mathrm{occ}^{T}.
\end{equation}
Similarly, taking the transpose of the generalized eigenvalue equation
and using the symmetry of \(\mat F\) and \(\bm\varepsilon_\mathrm{occ}\)
gives
\begin{equation}
  \mat D\mat F
  =
  \mat C_\mathrm{occ}\bm\varepsilon_\mathrm{occ}
  \mat C_\mathrm{occ}^{T}\mat S .
\end{equation}
It follows immediately that
\begin{equation}
  \mat F\mat D\mat S
  =
  \mat S\mat C_\mathrm{occ}\bm\varepsilon_\mathrm{occ}
  \mat C_\mathrm{occ}^{T}\mat S
  =
  \mat S\mat D\mat F .
\end{equation}
Thus
\begin{equation}
  \mat F\mat D\mat S-\mat S\mat D\mat F=0,
\end{equation}
which is Eq.~\eqref{eq:hf-ao}.  The density-matrix stationarity condition and
the Roothaan--Hall equation are therefore equivalent descriptions of the same
Hartree--Fock solution: the former expresses invariance of the occupied
subspace, while the latter chooses an explicit orthonormal basis within that
subspace.

\subsection{Summary}
The derivation proceeds entirely within the second-quantization AO formalism:
\begin{enumerate}
  \item The density matrix \(\mat D\) (equivalently, \(\bm\Delta\)) characterizes the Hartree--Fock state.
  \item The idempotency condition \(\mat D\mat S\mat D = \mat D\) constrains the state to be a single Slater determinant.
  \item The exponential parametrization \(\tilde{\mat D} = e^{\ii\bm\kappa\mat S}\mat D\,e^{-\ii\mat S\bm\kappa}\) generates all allowed variations while preserving idempotency and particle number.
  \item Stationarity of the energy with respect to \(\bm\kappa\) yields the Hartree--Fock equation \(\mat F\mat D\mat S - \mat S\mat D\mat F = 0\).
\end{enumerate}


\begin{thebibliography}{99}

\bibitem{Kjaergaard2008}
T.~Kjærgaard, P.~Jørgensen, J.~Olsen, S.~Coriani, and T.~Helgaker,
``Hartree--Fock and Kohn--Sham time-dependent response theory in a
second-quantization atomic-orbital formalism suitable for linear scaling,''
J.~Chem.~Phys.~\textbf{129}, 054106 (2008).

\bibitem{Larsen2000}
H.~Larsen, P.~Jørgensen, J.~Olsen, and T.~Helgaker,
``Hartree--Fock and Kohn--Sham atomic-orbital based time-dependent response theory,''
J.~Chem.~Phys.~\textbf{113}, 8908 (2000).

\bibitem{Roothaan1951}
C.~C.~J.~Roothaan,
``New developments in molecular orbital theory,''
Rev.~Mod.~Phys.~\textbf{23}, 69--89 (1951).

\bibitem{Hall1951}
G.~G.~Hall,
``The molecular orbital theory of chemical valency. VIII. A method of calculating ionization potentials,''
Proc.~R.~Soc.~Lond.~A \textbf{205}, 541--552 (1951).

\bibitem{Lowdin1955}
P.-O.~Löwdin,
``Quantum theory of many-particle systems. I. Physical interpretations by means of density matrices, natural spin-orbitals, and convergence problems in the method of configurational interaction,''
Phys.~Rev.~\textbf{97}, 1474--1489 (1955).

\bibitem{McWeeny1960}
R.~McWeeny,
``Some recent advances in density matrix theory,''
Rev.~Mod.~Phys.~\textbf{32}, 335--369 (1960).

\bibitem{Helgaker2000}
T.~Helgaker, P.~Jørgensen, and J.~Olsen,
\textit{Molecular Electronic-Structure Theory}
(Wiley, Chichester, 2000).

\bibitem{SzaboOstlund}
A.~Szabo and N.~S.~Ostlund,
\textit{Modern Quantum Chemistry: Introduction to Advanced Electronic Structure Theory}
(Dover, Mineola, 1996).

\end{thebibliography}
\end{document}